\documentclass[twoside,11pt]{article}

\usepackage[cp1251]{inputenc}
\usepackage{graphicx}

\usepackage{amssymb}
\usepackage{amsmath}
\usepackage{amsfonts}
\usepackage{wasysym}

\begin{document}           % End of preamble and beginning of text.

\title{\Large 
Low-velocity cosmic strings in accelerating universe
}
\author{V.E. Kuzmichev, V.V. Kuzmichev\\[0.5cm]
\itshape Bogolyubov Institute for Theoretical Physics,\\
\itshape National Academy of Sciences of Ukraine, Kiev, 03680 Ukraine}

\date{}

\maketitle

\begin{abstract}
The standard cosmological model supposes that the dominant matter component changes
in the course of the evolution of the universe.
We study the homogeneous and isotropic universe with non-zero cosmological constant
in the epoch when the dominant matter component has a form of a gas of low-velocity cosmic strings.
It is shown that after the scale transformation of the time variable such a model and
the standard model of a spatially flat universe filled with pressure-free matter provide the equivalent descriptions of
cosmological parameters as functions of time at equal values of the cosmological constant. 
The exception is the behavior of the deceleration parameter in the early universe. 
Pressure-free matter can obtain the properties of a gas of low-velocity cosmic strings
in the epoch when the global geometry and total amount of matter 
in the universe as a whole obey an additional constraint. 
This constraint follows from the quantum geometrodynamical approach in the semiclassical approximation. 
In terms of general relativity, its effective contribution to the field equations 
can be linked to the evolution in time of the equation of state of matter caused by
the processes of redistribution of energy between matter components. 
\end{abstract}


\section{Introduction}

The standard $\Lambda$CDM model (see, e.g., the reviews \cite{PDG,Rio})
gives the satisfactory description of the most
of the present cosmological data under the assumption of the existence of 
an antigravitating medium named dark energy 
as the largest constituent of mass-energy in the universe.  
At the same time, it is believed that a high level of fine-tuning is required
in this model.
Even if the smallness of cosmological constant and ``coincidence problem'' 
(an almost equal contribution of matter and dark energy to the total energy budget of the universe at 
the present era) are not problems in themselves \cite{Bia},  nevertheless 
it should not be ignored that there were some indications that specific cosmological observations
differed from the predictions of the $\Lambda$CDM model at statistically significant level \cite{Pe}.

The $\Lambda$CDM model based on general relativity allows extensions by incorporating new
elements in its scheme.
For example, one of such possibilities is the introduction of the quintessence field, which changes over time and is described by some dynamic equation, instead of the cosmological constant. 
Another opportunity may be the model in which, alternatively, the gravitating matter component 
undergoes modification, regardless of the vacuum component of the
energy density being constant or varying with time. Such a modification may be made relying on
fundamental physical laws which concern the properties of matter.

In FRW cosmology, the time evolution of the energy density $\rho (t)$ is determined by the equation
\begin{equation}\label{0}
   \dot{\rho} + 3 \frac{\dot{R}}{R} (\rho + p) = 0,
\end{equation}
where $R(t)$ is the cosmological scale factor, $p$ is the isotropic pressure, and the dot 
designates the derivative with respect to the proper time $t$. For the equation of state in the form
$p = w \rho$, the solution of this equation vanishing at infinity can be written as
$\rho = \mu R^{-3(1 + w)}$, where $w$ and $\mu$ are constants. Introducing the effective
mass $M_{eff}$ contained in the volume $\sim R^{3}$ by the relation $M_{eff} \sim \rho R^{3}$,
we have $M_{eff} \sim \mu R^{-3w}$. For the special case $w = 0$, it gives $M_{eff} \sim \mu = const$
which corresponds to pressure-free matter (dust). 
For $w = - \frac{1}{3}$, the effective mass is proportional to the
scale factor, $M_{eff} \sim \mu R$. In this case the energy density $\rho \sim R^{-2}$ and it describes
so-called K-matter \cite{Kol}.  The matter with such energy density and equation of state can be interpreted as a perfect gas of low-velocity cosmic strings \cite{KT}. 

In this paper we study the model of the homogeneous and isotropic 
universe with non-zero cosmological constant filled with a perfect gas of low-velocity cosmic strings.
Throughout the paper, we will refer to this model as the $\Lambda$CS model. 
It is shown that pressure-free matter can obtain the properties of a gas of low-velocity cosmic strings, 
if, in addition to the field equations, there exists a complementary constraint between the global geometry and 
total amount of matter in the universe as a whole.
We show that this constraint between the cosmological parameters, which takes the form of the geometry-mass relation,
can be obtained in the quantum geometrodynamical approach. 
In terms of general relativity, its effective contribution to the field equations 
can be linked to the evolution in time of the equation of state of matter caused by
the processes of redistribution of energy between matter components. This is demonstrated in the model
in which two-component perfect fluid serves as a surrogate for matter in the universe.

We found the exact solutions of the Einstein equations for the $\Lambda$CS model. 
It is demonstrated that this model is equivalent to the open de Sitter model. In the limit of zero 
cosmological constant, the corresponding universe evolves as a Milne universe characterized by
the linear dependence of the scale factor on time, but in contrast to it, 
such a universe contains matter with nonzero energy density in the form of a perfect gas of low-velocity cosmic strings. 
The Whitrow-Randall equation \cite{Whi} which establishes the invariance of the dimensionless product
$G \rho t^{2}$ is re-derived.
We make a comparison of the standard $\Lambda$CDM and $\Lambda$CS models.
It turns out that after the scale transformation $t \to \frac{3}{2} t$ of the time variable of the 
$\Lambda$CS model, these models provide the equivalent descriptions of cosmological parameters
as functions of time at equal values of the cosmological constant.
The exception is the behavior of the deceleration parameter in the early universe.
But for the present day and in the future
it would be more difficult to recognize whether one is dealing with the $\Lambda$CDM or $\Lambda$CS universe.

\section{Quantum roots of the geometry-mass relation}

It is well known that quantum theory adequately 
describes properties of various physical systems. Its universal validity demands that the universe as a 
whole must obey quantum laws as well, so that quantum effects are important at least in the early era. 
Since quantum effects are not a priori restricted to certain scales, then one should not conclude in advance 
that they cannot have any impact on processes at scales larger than Planckian 
(more detailed arguments can be found, e.g., in Refs. \cite{Mes}).

Quantum theory for a homogeneous and isotropic universe can be constructed on the basis of a 
Hamiltonian formalism with the use of material reference system as a dynamical system \cite{Ku1,Ku2}. 
Defining the time parameter or the ``clock'' variable, it is possible to pass from 
the Wheeler–DeWitt equation to the Schr\"odinger-type equation.
The similar equations containing a time variable defined by means of coordinate condition were 
considered by a number of authors under the quantization of the FRW universe 
(see, e.g., Refs. \cite{Lun}).
Using the Schr\"odinger-type equation one can obtain equations of motion for the expectation values of 
a scale factor and its conjugate momenta. These equations pass into the equations of general relativity 
when the dispersion around the expectation values for a scale factor, matter fields and their conjugate 
momenta can be neglected.

Such a quantum theory predicts that the following relation must hold
for the expectation value of the scale factor $R$ in the state $| M \rangle$ which describes the universe with the definite total amount of mass $M$ much larger than Planck mass, $M \gg M_{P}$,
\begin{equation}\label{1}
\frac{\langle M | R | M \rangle}{\langle M | M \rangle} = G M
\end{equation}
(in units $c = 1$; for details, see Refs. \cite{Ku2}), where $G$ is the Newtonian gravitational constant. 

The equation (\ref{1}) determines the mass $M$ through the expectation value of the scale factor $R$
at every instant of time. The state vector of isotropic universe is a superposition of all possible $| M \rangle$ - states which are not orthogonal between themselves, so that the inner product $\langle M_{1} |M_{2} \rangle \neq 0$, and the universe can transit spontaneously from the state with the mass $M_{1}$ and radius $R_{1} = \langle M_{1} | R | M_{1} \rangle / \langle M_{1} |M_{1} \rangle$
to the state with the mass $M_{2} \neq M_{1}$ and radius $R_{2} = \langle M_{2} | R | M_{2} \rangle / \langle M_{2} |M_{2} \rangle$
with nonzero probability $P(1\rightarrow 2) = |\langle M_{1} |M_{2} \rangle|^{2}$. For example, the probability of transition of the universe from the ground state (with respect to gravitational field) to any other state obeys the Poisson distribution with the mean number of occurrences $n = \frac{1}{2}(M_{2} - M_{1})^{2}$ (for more details, see Refs. \cite{Ku2}). Thus $R_{1} \rightarrow R_{2}$, when $M_{1} \rightarrow M_{2}$. 

In classical limit, it appears to be possible to pass from the expectation value of $R$ to the classical value of the scale factor $R(t)$ which evolves in time in accordance with the Einstein equations for the FRW universe
\begin{eqnarray}\label{2}
\dot{R}^{2} & = & \frac{8\pi G}{3}\,\rho R^{2} + \frac{\Lambda}{3}\,R^{2} - k, \nonumber \\
\ddot{R} & = & - \frac{4\pi G}{3}(\rho + 3p)R + \frac{\Lambda}{3}\,R,
\end{eqnarray}
where 
\begin{equation}\label{3}
    \rho = \frac{M}{(4\pi /3) R^{3}} 
\end{equation}
is the energy density of matter with the total mass $M$ in the equivalent flat-space volume $(4\pi /3) R^{3}$ which includes both the mass
of substance and mass equivalent of radiation energy, $\Lambda$ is the cosmological constant,
\begin{equation}\label{4}
    p = - \rho - \frac{R}{3}\,\frac{d\rho}{dR}
\end{equation}
is the isotropic pressure, and $k = +1,0,-1$ for spatially closed, flat or open models. 
In semi-classical limit, the relation (\ref{1}) takes the form
\begin{equation}\label{5}
    R = G M.
\end{equation}
It gives an additional constraint between the global geometry and total amount of matter in the universe as a whole.
The geometry-mass relation (\ref{5}) connects the values of $M$ and $R$ taken
at the same instant of time. It is valid for the present-day universe. 
Really, the radius of its observed part is estimated as $R_{0} \sim 10^{28}$ cm, 
the mass-energy is $M_{0} \sim 10^{56}$ g, and the mean energy density equals to $\rho_{0} \sim 10^{-29}\mbox{g cm}^{-3}$. 
It means that nowadays $\rho_{0} \sim 3 (4 \pi G R_{0}^{2})^{-1}$. Then from the definition of energy 
density $\rho_{0} = 3 M_{0} (4 \pi R_{0}^{3})^{-1}$, 
it follows that the relation $R_{0} \sim GM_{0}$ must hold. 
It is notable that for the values $R = L_{P}$, $M = M_{P}$, where $L_{P}$ is the Planck length, the equation (\ref{5}) reduces to identity.

The physical meaning of the relation (\ref{5}) will be discussed in Sect.~5. Here we remark only that since it is valid at
least at late times $t \sim t_{0}$, where $t_{0}$ is the age of the universe, then the theory which includes the geometry-mass relation
(\ref{5}) may be used for the description of the evolution of the universe on the interval $t = t_{0} \mp \Delta t$, $\Delta t / t_{0} \ll 1$.

\section{$\Lambda$CS model}

If one supposes that the values $R$ and $M$ in Eq. (\ref{5}) are constant, then the FRW universe described by the equations (\ref{2})
transforms into the static Einstein universe \cite{Tol}. Let us consider the more general case assuming that the relation (\ref{5}) 
is valid for some time interval and can be regarded as a constraint added to the classical field equations (\ref{2}).
Then the energy density of matter (\ref{3}) takes the form of energy density of a gas of low-velocity 
cosmic strings or K-matter \cite{Kol,KT} with the corresponding equation of state,
\begin{equation}\label{6}
   \rho = \frac{3}{G} \frac{1}{4 \pi R^{2}}, \quad p = - \frac{1}{3} \rho.
\end{equation}
However, in this approach it does not mean that the universe is string-dominated in the usual sense.
The energy density and pressure in the form (\ref{6}) arise as an effect of an additional constraint between the global geometry and the total amount of matter in the universe as a whole.

Let us consider the model of the universe with cosmological constant $\Lambda_{s}$ and the matter density and pressure as in Eq. (\ref{6}). The field equations are reduced to the form
\begin{equation}\label{7}
\dot{R}^{2} = \frac{\Lambda_{s}}{3}\,R^{2} + (2 - k), \qquad
\ddot{R} = \frac{\Lambda_{s}}{3}\,R.
\end{equation}
Their solution is
\begin{equation}\label{8}
    R_{s} (t) = \sqrt{\frac{6}{\Lambda_{s}} \left(1 - \frac{1}{2}k \right)} 
    \sinh \left(\sqrt{\frac{\Lambda_{s}}{3}} t \right), \ R_{s} (0) = 0.
\end{equation}
Here and below the subscript $s$ refers to the $\Lambda$CS model.
Expansion of this solution for small $|\frac{\Lambda_{s}}{3} t^{2}|$ yields
\begin{equation}\label{9}
    R_{s} (t) = \sqrt{2- k} t \left[1 + \frac{1}{6} \left(\frac{\Lambda_{s}}{3} t\right)^{2} + \ldots \right].
\end{equation}
The Hubble expansion rate does not depend on the type of spatial curvature (the value of $k$) and
it is described by the expression
\begin{equation}\label{10}
    H_{s} (t) = \frac{\dot{R_{s}}}{R_{s}} = 
    \sqrt{\frac{\Lambda_{s}}{3}} \coth \left(\sqrt{\frac{\Lambda_{s}}{3}} t \right).
\end{equation}
The expansion of $H_{s}(t)$ in the same limit as above has a form
\begin{equation}\label{11}
    H_{s} t = 1 + \frac{1}{3} \left(\frac{\Lambda_{s}}{3}\right) t^{2} - 
    \frac{1}{45} \left(\frac{\Lambda_{s}}{3}\right)^{2} t^{4} + \ldots
\end{equation}

In general case, the Hubble expansion rate $H$ 
is the function of time and the corresponding critical energy density is
$\rho_{cr} (t) = \frac{3 H^{2} (t)}{8 \pi G}$. Then the variation in time of the 
vacuum energy density parameter 
$\Omega_{\Lambda} (t) \equiv \frac{\rho_{\Lambda}}{\rho_{cr} (t)}$,
where $\rho_{\Lambda} = \frac{\Lambda}{8 \pi G}$, for the $\Lambda$CS model is given by
\begin{equation}\label{1111}
    \Omega_{\Lambda s} (t) = \tanh^{2}\left(\sqrt{\frac{\Lambda_{s}}{3}}t\right).
\end{equation}

The matter energy density parameter $\Omega_{M} (t) = \frac{\rho}{\rho_{cr} (t)}$ for a
spatially flat universe filled with a gas of low-velocity cosmic strings is equal to
\begin{equation}\label{1112}
    \Omega_{M s} (t) = \cosh^{-2}\left(\sqrt{\frac{\Lambda_{s}}{3}}t\right).
\end{equation}

Setting $\Lambda_{s} = 3\, \Omega_{\Lambda 1} H_{s}^{2}(t_{1})$, 
where $t_{1}$ is some fixed instant of time,
$\Omega_{\Lambda 1}$ is the vacuum energy density parameter $\Omega_{\Lambda s}$ at $t = t_{1}$,
we find
\begin{equation}\label{111}
    H_{s} (t_{1})\, t_{1} = \frac{1}{\sqrt{\Omega_{\Lambda 1}}} \,\mbox{arctanh} \sqrt{\Omega_{\Lambda 1}}.
\end{equation}

The deceleration parameter $q(t) = - \frac{\ddot{R}}{R H^{2} (t)}$ is equal to
\begin{equation}\label{112}
   q_{s}(t) = - \Omega_{\Lambda s} (t).
\end{equation}
At an instant of time $t = t_{1}$ we obtain $q_{s}(t_{1}) = - \Omega_{\Lambda 1}$.

If $\Lambda_{s} \neq 0$, the expressions for the scale factor (\ref{8}) and the Hubble expansion rate 
(\ref{10}) are equivalent to the respective expressions for the de Sitter model of the universe with 
$k = -1$.

In the limiting case $\Lambda_{s} = 0$ it appears that
\begin{equation}\label{12}
    R_{s} (t) = \sqrt{2- k} t, \qquad H_{s} t = 1.
\end{equation}
In the model, where the scale factor depends on time linearly, the age of the universe and
the Hubble expansion rate depend on the redshift $z$ according to the simple laws
\begin{equation}\label{15}
    t (z) = \frac{1}{(1 + z) H_{s} (0)}, \qquad H_{s} (z) = H_{s} (0) (1 + z).
\end{equation}
Taking the present expansion rate measured by Hubble Space Telescope observations,
$H(0) = 73.8 \pm 2.4$ km s$^{-1}$ Mpc$^{-1}$ \cite{Rie}, as $H_{s}(0)$, we find that
the age of the universe appears to be equal $t(0) = 13.26 \pm 0.43$ Gyr. 
This value does not differ drastically from the value predicted by
the WMAP 7-year data \cite{Lar} for the $\Lambda$CDM model, and it lies within the expected limit of 12 to 14 Gyr.

The solution (\ref{12}) formally coincides with the solution of Milne model of open universe ($k = -1$), 
$R (t) \sim t$.
But in contrast to the Milne model, where the energy density of matter vanishes, $\rho = 0$, in
the case under consideration the energy density of matter is nonzero,
\begin{equation}\label{13}
    \rho = \frac{3 H^{2}}{4 \pi G (2 - k)}.
\end{equation}
For a spatially flat universe with zero cosmological constant this density equals to the critical density, 
$\rho = \rho_{cr}$, whereas for a spatially closed universe filled 
with a gas of low-velocity cosmic strings the density is $\rho = 2\rho_{cr}$, and for a spatially open 
universe we have $\rho = \frac{2}{3} \rho_{cr}$.

It should be noted that
the Milne model cannot be correct near the point of initial cosmological singularity, $t = 0$, since
in this limit the energy density of matter tends to infinity and gravity cannot be neglected. There was
an attempt to preserve the linear dependence of a scale factor on time and 
get rid of this shortcoming of the Milne model by consideration of the model (called ``Dirac - Milne'' 
universe by analogy of sea of positive and negative energy states proposed by Dirac), 
in which the universe contains equal quantities of matter
with positive and negative gravitational masses
\cite{Ben}. 

The equation (\ref{13}) can be rewritten in the Whitrow-Randall form \cite{Whi}, 
\begin{equation}\label{131}
  G \rho t^{2} = \frac{3}{4 \pi} \frac{1}{n},
\end{equation}
which shows that $G \rho t^{2}$ is an invariant determined by the parameter $n = 2 - k$ characterizing the geometry of the universe.

Introducing a dimensionless parameter $K$ as in the model of K-matter,
\begin{equation}\label{14}
    K \equiv \frac{8 \pi G}{3} \rho R^{2},
\end{equation}
and using (\ref{6}), one finds that $K = 2$. This value agrees with the observational constraints on 
the parameter $K$ obtained by Kolb \cite{Kol} and Gott and Rees \cite{Got}.

\section{Comparison with the standard cosmological model}

The relations obtained in Sect.~3 for the cosmological parameters of the $\Lambda$CS model of
the universe can be compared with the
corresponding expressions for the standard cosmological model. First of all, it is helpful to rewrite
the equation for $\dot{R}$ from Eq. (\ref{2}) in the form
\begin{equation}\label{141}
   \dot{R}^{2} = \frac{\Lambda}{3}\,R^{2} + (2 - k) + \zeta(R),
\end{equation}
where the function $\zeta(R)$ is defined as
\begin{equation}\label{142}
   \zeta(R) = \frac{2}{R} (GM - R).
\end{equation}
Comparing Eqs. (\ref{7}) and (\ref{141}) we find that the function $\zeta(R)$
describes the difference between the model which takes into account the geometry-mass relation (\ref{5}) and
the model without regard for it.

For a spatially flat universe ($k = 0$) filled with pressure-free matter ($p = 0$, $M = \mbox{const}$) 
the solution of Eq. (\ref{141}) has a form (cf. Ref. \cite{Gr})
\begin{equation}\label{143}
   R (t) = \left (\frac{6}{\Lambda} GM \right)^{1/3} 
   \sinh^{2/3} \left(\frac{3}{2} \sqrt{\frac{\Lambda}{3}}\, t \right), \ R (0) = 0.
\end{equation}

From Eqs. (\ref{8}) and (\ref{143}) it follows that the scale factors of both models can be connected by
the relation
\begin{equation}\label{149}
   R(t) = \left(\frac{G M}{1 - \frac{1}{2}k}\right)^{1/3} R_{s}^{2/3}\left(\frac{3}{2} t\right ) \quad \mbox{at}
   \quad \Lambda = \Lambda_{s}.
\end{equation}
This expression establishes the rule of recalculation of the scale factor from the $\Lambda$CS model 
with arbitrary spatial curvature to the spatially flat standard model. This connection between two 
models becomes more clear if one takes into account in the solution (\ref{143}) the relation (\ref{5})
in its weaker form, namely, assuming that it is valid at some fixed instant of time $t_{0}$ only.
As we have already mentioned above, this relation is realized in the present-day universe.
Setting $R(t_{0}) = G M(t_{0})$, where $M(t_{0})$ is the mass of matter in the universe
with the ``radius'' $R(t_{0})$, we have
\begin{equation}\label{150}
   R (t_{0}) = \sqrt{\frac{6}{\Lambda}}
   \sinh \left(\frac{3}{2} \sqrt{\frac{\Lambda}{3}}\, t_{0} \right).
\end{equation}
Then the relation (\ref{149}) at fixed instant of time takes the form
\begin{equation}\label{151}
   R(t_{0}) = \left(1 - \frac{1}{2}k \right)^{-1/2} R_{s} \left(\frac{3}{2} t_{0}\right ) \quad \mbox{at}
   \quad \Lambda = \Lambda_{s}.
\end{equation}

The Hubble expansion rate in the standard model is
\begin{equation}\label{144}
    H (t) = \frac{\dot{R}}{R} = \sqrt{\frac{\Lambda}{3}} \coth \left(\frac{3}{2}\sqrt{\frac{\Lambda}{3}} t \right).
\end{equation}
Comparing Eq. (\ref{144}) with Eq. (\ref{10}), we find that the Hubble expansion rates calculated for 
both models are related between themselves by the simple relation
\begin{equation}\label{145}
    H (t) = H_{s}\left(\frac{3}{2} t \right) \quad \mbox{at} \quad \Lambda = \Lambda_{s}.
\end{equation}
The expansion of $H (t)$ for small $|\frac{\Lambda}{3} t^{2}|$ can be obtained 
from Eq. (\ref{11}) by the substitution of $\frac{3}{2} t$ for $t$ and $\Lambda$ for $\Lambda_{s}$.
That expansion of $H (t)$ reproduces the familiar expression $H = \frac{2}{3 t}$ for
$\Lambda = 0$ in contrast to $H_{s} = \frac{1}{t}$ for the $\Lambda$CS model with $\Lambda_{s} = 0$.

The Hubble expansion rates as the functions of dimensionless time parameter 
$t/t_{0}$, where $t_{0}$ is the age of the universe in the $\Lambda$CDM model, are plotted in Fig.~1. 
It is supposed that the cosmological constant in both models is the same. 
The WMAP 7-year data \cite{Lar} for the present-day cosmological parameters are used.
As we can see, 
the value of $H(t)$ at $t = t_{0}$ coincides with the value of $H_{s}(t)$ at $t =  \frac{3}{2}\,t_{0}$.
%Fig.~1
\begin{figure*}% figure* for wide figure, [h] [!] to change the placement
\includegraphics[width=10cm]{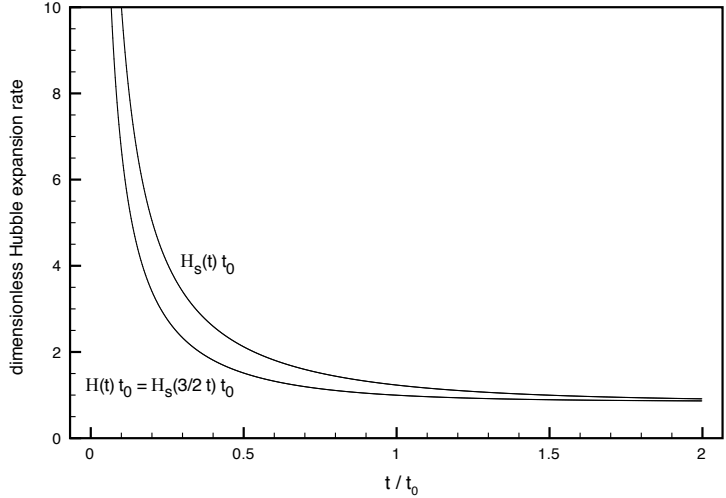}
\caption{The Hubble expansion rates (\ref{10}) and (\ref{144}) as functions of time. 
The value of the parameter
$\sqrt{\frac{\Lambda}{3}} t_{0} = 0.855$ following from the WMAP 7-year data \cite{Lar} is used.
It is assumed that $\Lambda = \Lambda_{s}$.} %\label{fig:1}
\end{figure*} 

Taking into account Eqs. (\ref{145}), (\ref{1111}) and (\ref{1112}), we find the relations
\begin{eqnarray}\label{1451}
    \Omega _{\Lambda}(t) & = & \Omega _{\Lambda s}\left(\frac{3}{2} t \right), \nonumber \\
    \Omega _{M}(t) & = & \Omega _{M s}\left(\frac{3}{2} t \right)
    \quad \mbox{at} \quad \Lambda = \Lambda_{s},
\end{eqnarray}
where $\Omega _{\Lambda}(t)$ and $\Omega _{M}(t)$ are the vacuum and matter energy density
parameters of the standard model.

The energy density parameters of the vacuum and matter as functions of time are depicted in Fig.~2.
It is assumed that the universe is spatially flat. As for the Hubble parameter, the
$\Lambda$CS model reproduces the results of the standard model after the time transformation
$t \to \frac{3}{2} t$.
%Fig.~2
\begin{figure*}% figure* for wide figure, [h] [!] to change the placement
\includegraphics[width=10cm]{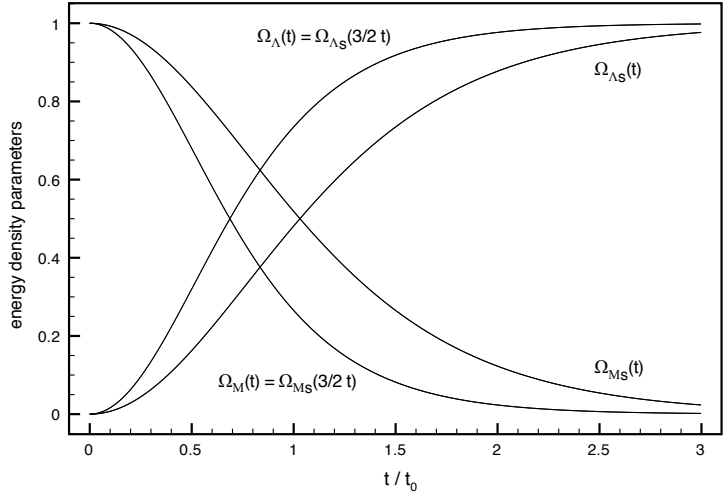}
\caption{The energy density parameters of the vacuum and matter (\ref{1111}), (\ref{1112}), and
(\ref{1451}) as functions of time in units of $t_{0}$. 
The WMAP 7-year data are used (see the caption of the Fig.~1).} %\label{fig:2}
\end{figure*} 

Introducing the vacuum energy density parameter $ \Omega_{\Lambda 0}$ at a fixed instant of time
$t = t_{0}$ by the relation $\Lambda = 3\, \Omega_{\Lambda 0} H^{2}(t_{0})$, we get
\begin{equation}\label{146}
    H (t_{0})\, t_{0} = \frac{2}{3}\, \frac{1}{\sqrt{\Omega_{\Lambda 0}}} \,
    \mbox{arctanh} \sqrt{\Omega_{\Lambda 0}}.
\end{equation}
From the comparison of Eq. (\ref{146}) with Eq. (\ref{111}) it follows that at equal Hubble expansion rates 
according to Eq. (\ref{145}) and equal contributions of the vacuum energy densities 
into the matter-energy budget of the universe, 
$\Omega_{\Lambda 0} = \Omega_{\Lambda 1}$, the parameter $t_{1} = \frac{3}{2} t_{0}$.
It means that if one defines $t_{0}$ and  $t_{1}$ as the ages of the universe in both models 
under consideration, then the age $t_{1}$ for the $\Lambda$CS model will be 1.5 times greater 
than that for the standard model.

The last equation establishes the correspondence between the parameters 
$H (t_{0})$, $t_{0}$, and $\Omega_{\Lambda 0}$. 
Substituting the WMAP 7-year data \cite{Lar} for the age of the universe $t_{0} = 13.75 \pm 0.13$ Gyr,  
the Hubble parameter $H (t_{0}) = 71.0 \pm 2.5$ km s$^{-1}$ Mpc$^{-1}$, and the dark energy density parameter $\Omega_{\Lambda 0} = 0.734 \pm 0.029$, which corresponds to the cosmological constant $\Lambda = (1.302 \pm 0.143) \times 10^{-56}$ cm$^{-2}$, into Eq. (\ref{146})
one finds a consistent result for our universe: $H (t_{0}) t_{0} = 0.998 \pm 0.045$.

The deceleration parameter $q$ is equal to
\begin{equation}\label{147}
   q (t) = \frac{1}{2} \left [1 - 3 \tanh^{2}\left(\frac{3}{2}\, \sqrt{\frac{\Lambda}{3}}t\right)
   \right ].
\end{equation}
At an instant of time $t = t_{0}$ it gives
\begin{equation}\label{148}
   q (t_{0}) = \frac{1}{2} \left [1 - 3 \Omega_{\Lambda 0} \right ].
\end{equation}
Comparing Eqs. (\ref{112}) and (\ref{147}), we find that both expressions for the 
deceleration parameter have the same limit at $t \rightarrow \infty$, $q(t) \rightarrow -1$ and
$q_{s}(t) \rightarrow -1$, but they have different values at $t = 0$, $q(0) = \frac{1}{2}$ and 
$q_{s}(0) = 0$.
From the condition (\ref{145}) valid at $\Lambda = \Lambda_{s}$, we have 
$\Omega_{\Lambda 0} = \Omega_{\Lambda 1}$, and the expression (\ref{148}) can be rewritten as
\begin{equation}\label{1481}
   q(t_{0}) = \frac{1}{2} \left [1 + 3 q_{s}\left(\frac{3}{2} t_{0} \right) \right ].
\end{equation}

In Fig.~3 it is shown the time dependence of the deceleration parameters $q (t)$ and $q_{s} (t)$ for the standard and $\Lambda$CS models. The function $q_{s}\left(\frac{3}{2} t\right)$ 
with the argument multiplied by $\frac{3}{2}$ is plotted for comparison. 
For $\frac{t}{t_{0}} \gtrsim 2.5$ the curves $q(t)$ and $q_{s}\left(\frac{3}{2} t\right)$ practically coincide. 
The both models  predict the accelerating expansion of the universe at $t = t_{0}$ and give the close 
values of the deceleration parameter, $q(t_{0}) = - 0.601$ and 
$q_{s}\left(\frac{3}{2} t_{0}\right) = - 0.734$. 
They lead to the same limit at $t \rightarrow \infty$, $q(t) \rightarrow -1$ and $q_{s}(t) \rightarrow -1$.  
But in the region $\frac{t}{t_{0}} < 1$ 
the behaviors of the functions differ drastically. The standard model predicts that at 
$\frac{t}{t_{0}} < 0.5$ the universe was decelerating. On the contrary, the $\Lambda$CS model 
describes the always accelerating universe with non-zero cosmological constant and containing matter in the form of a perfect gas of low-velocity cosmic strings.

%Fig.~3
\begin{figure*}% figure* for wide figure, [h] [!] to change the placement
\includegraphics[width=10cm]{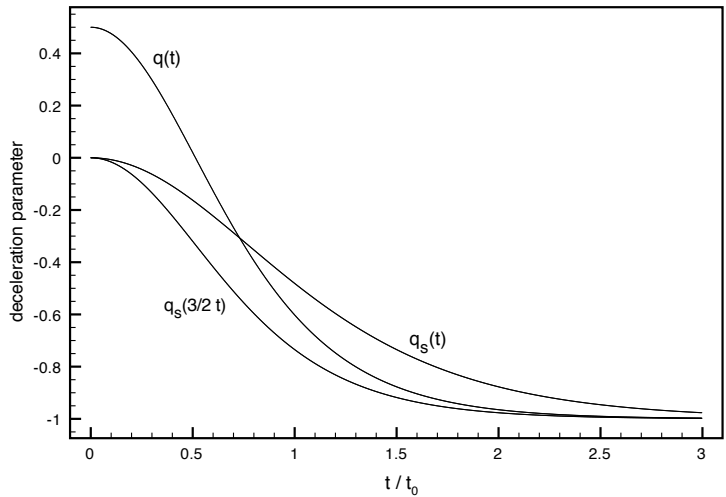}
\caption{The deceleration parameters (\ref{112}) and (\ref{147}) as functions of time in units of $t_{0}$. 
The WMAP 7-year data are used (see the caption of the Fig.~1)} %\label{fig:3}
\end{figure*} 

The fact that predictions concerning the deceleration parameter of both models differ for the
time interval $0 < t < \frac{1}{2} t_{0}$ is not surprising. The $\Lambda$CS model in the form under consideration
does not claim to provide a satisfactory description of the universe at all times. 
This model is inapplicable to describe the 
expansion of the universe at times when the geometry-mass relation (\ref{5}) has no impact on the 
properties of matter.

At the same time, we can conclude that for the present-day and future
universe it would be more difficult to distinguish between the $\Lambda$CDM and $\Lambda$CS models.
It looks like the universe at some instant of time becomes dominated by matter in the form of 
a perfect gas of low-velocity cosmic strings. 
The reason of such a transformation of matter
is different from that which leads to the formation of macroscopic topological defects in the form of strings
in the early universe where they are caused by phase transitions (see Ref. \cite{Vil} and references therein). In the later universe
the energy density and pressure in the form as for a perfect gas of low-velocity cosmic strings may arise as an effect of an additional constraint between the global geometry and the total amount of matter in the universe as a whole. 

Thus, the $\Lambda$CDM and $\Lambda$CS  
models lead to similar predictions on cosmological parameters, if the time variable of the 
$\Lambda$CS model is subject to scale transformation $t \to \frac{3}{2} t$. After this transformation,
the Hubble expansion rate and the energy density parameters of matter and vacuum components of the universe, 
calculated for all instants of time, take equal values in both models. 

\section{Discussion}

Let us consider the possible physical interpretation of the geometry-mass relation (\ref{5}). First of all, we point out that 
the similar equality between the mass and ``radius'' of the universe was obtained by Whitrow and Randall \cite{Whi}
(see Eq. (\ref{131}) for $k = 0$). 
Such a relation is also valid for the Einstein universe filled with the pressure-free matter (see, e.g., Ref. \cite{Tol})
and for the steady-state cosmology \cite{Row}.

Further on, it should be noted that the relation (\ref{5}) has a form of Sciama's inertial force law $M \sim G^{-1} R$, 
where $M$ and $R$ are appropriate measures of mass and radius of the observed part of the universe \cite{Sc1,Sc2}.
Despite its simplified character, Sciama's linearized theory gives a specific mathematical relation between 
the parameters which characterize the energy density and geometry of the universe and corresponds to 
one of the realizations of Mach's principle \cite{Mach,Bon,Bar}.

If one assumes that Mach's principle is a fundamental law of nature, it must be implemented into the 
classical field equations. One point of view is that Einstein's field equations need not to be modified,
while Mach's principle should be considered as an additional condition. Such an approach was chosen
by Wheeler who proposed to understand Mach's principle as a selection rule (boundary condition) of
the solutions of the field equations \cite{Whe}. The Brans-Dicke theory 
uses another way in which the field equations are generalized to become Machian \cite{Bra,Hel}.

Since the scale factor $R$ obeys the equations (\ref{2}), then
the mass $M$, generally speaking, must evolve in time. It means that if the gravitational constant
$G$ and velocity of light $c$ are both constant, the mass of matter in the universe must change
proportionally to the scale factor, $M \sim R(t)$, at the time interval, where the relation (\ref{5})
holds.

In some cosmological models the natural constants $G$ or $c$ are suppose to change with time.
For example, according to Dirac’s large number hypothesis, the Newtonian constant $G$ must depend on time as
$G \sim t^{-1}$ and $R \sim t^{1/3}$ \cite{Di1} or $G \sim t^{-1}$ and $R \sim t$ \cite{Di2}. 
Another example with varying $G$ is the Brans-Dicke theory where this quantity is related to the average value of some dynamical scalar field $\phi$ coupled to the mass density $\rho$ of the universe, $\langle \phi \rangle \approx G^{-1}$, where $\langle \phi \rangle \sim \rho R^{2}$ \cite{Bra,Wei}.
Also the models with varying speed of light were considered and applied to solve the horizon, flatness,  cosmological constant, and other cosmological problems (see, e.g., Refs. \cite{Pet}). On the other hand, there exist the observational and experimental bounds on the time variation of the fundamental constants (e.g., Ref. \cite{Chi}). 

The possible dependence of mass $M$ on time can be considered in terms of the fundamentally different approach which
deals with the matter creation processes in the context of the cosmological models \cite{Lim}. But currently the models with the irreversible 
creation of matter do not rely on sufficient observational evidence.

We shall use another approach which does not take into account the theoretical schemes mentioned above in this Section.
The relation (\ref{5}) follows from the quantum theory in semiclassical approximation. In terms of general relativity its
effective contribution into the field equations can be linked to the evolution in time of the equation of state of matter caused by
the processes of redistribution of energy between its components. 

Let us consider the model in which the equation of state parameter for matter,
\begin{equation}\label{200}
	 w(t) = \frac{p(t)}{\rho(t)},
\end{equation}
depends on time\footnote{Throughout this Section we shall assume that $\Lambda = 0$ for simplicity.}. 

In the context of hot big bang cosmology, the radiation-dominated universe with
the energy density $\rho \sim R^{-4}$ in the course of the expansion transforms into the non-relativistic matter-dominated universe with the energy density $\rho \sim R^{-3}$,
and the latter transits to a universe which looks like dominated by a perfect gas of low-velocity cosmic 
strings with $\rho \sim R^{-2}$ at later time.
In the radiation-dominated universe the number density of photons is $n_{\gamma} \sim R^{-3}$,
and the energy of every photon decreases, during the expansion of the universe,
as $m_{\gamma} \sim R^{-1}$ due to the cosmological redshift. 
As a result, the effective mass of the universe attributed to relativistic matter reduces as well,
$M_{eff} \sim m_{\gamma}  n_{\gamma}  R^{3} \sim R^{-1}$. Arguing in the same way, one finds that in 
the matter-dominated universe the effective mass is constant, $M_{eff} = const$, expressing the 
constancy of the sum of the masses of the bodies in the volume $\sim R^{3}$. In the universe 
which looks like dominated by a perfect gas of low-velocity cosmic strings, 
the effective mass of matter increases with the expansion of the universe,
$M_{eff} \sim R$, due to the redistribution of energy between the matter components. Thus, we have 
the following picture of changes of dominating matter content of the universe during its evolution in time: the mass of dominating matter component
in the expanding universe decreases inversely proportional to the ``radius'' $R$ of the universe in 
the radiation-dominated era, then the mass remains constant in the matter-dominated era, and finally it 
increases linearly with $R$ when a gas of
point particles (dust) transforms effectively into a perfect gas of low-velocity cosmic strings.
At the same time, the equation of state of dominating matter changes from the equation 
$p = \frac{1}{3}\rho$ to $p = - \frac{1}{3}\rho$ passing through the state $p = 0$.

According to the scenario being described here, we shall specify the parameter $w(t)$ in the form of antikink,
\begin{equation}\label{201}
	 w(t) = - \frac{1}{3}\,\tanh[\lambda (t - t_{0})],
\end{equation}
where $t_{0}$ is the instant of time in the neighborhood of which matter behaves as pressure-free dust ($t_{0}$ may be taken
close to the age of the universe), $\lambda$ is some parameter averaged in time which determines the rate of change of 
the equation of state.

The parameter (\ref{201}) can be justified in the model in which matter in the universe is described as a two-component perfect fluid
with the energy density $\rho = \rho_{1} + \rho_{2}$ and pressure $p = p_{1} + p_{2}$. We shall represent 
the energy conservation equation (\ref{0}) for every component in the form
\begin{equation}\label{202}
   \dot{\rho_{1}} + 3 H (\rho_{1} + p_{1}) = Q, \quad \dot{\rho_{2}} + 3 H (\rho_{2} + p_{2}) = -Q,
\end{equation}
where $Q$ describes the interaction between the components.

We shall assume that the equation of state for the component with the density $\rho_{1}$ changes in time from the
stiff Zel’dovich type equation $p_{1} = \rho_{1}$ to the vacuum type one  $p_{1} = - \rho_{1}$. The second component
is a pressure-free matter which has the density $\rho_{2} = 2 \rho_{1}$. Then $Q = 2 H p_{1}$ and $w = p_{1}/(3 \rho_{1})$,
while the set of equations (\ref{202}) reduces to one equation
\begin{equation}\label{203}
   \dot{\rho_{1}} + 3 H \left(\rho_{1} + \frac{1}{3} p_{1}\right) = 0.
\end{equation}
We look for the energy density $\rho_{1}$ in the form
\begin{equation}\label{204}
   \rho_{1} = \frac{\alpha(t)}{t^{2}},
\end{equation}
where $\alpha(t)$ is a slowly varying function on the interval $0 < t < \infty$. Then from the Einstein equation
for $\dot{R}$ and Eq. (\ref{203}), in the approximation $\dot{\alpha} \ll \lambda \alpha$, we find the dependence
of the total energy density $\rho$ and pressure $p$ on time,
\begin{eqnarray}\label{205}
   \rho (t) & = & \frac{3}{2 \pi G t^{2}}\,\frac{1}{[3 - \tanh (\lambda (t - t_{0}))]^{2}}, \nonumber \\
   p(t) & = & - \frac{1}{2 \pi G t^{2}}\,\frac{\tanh (\lambda (t - t_{0}))}{[3 - \tanh (\lambda (t - t_{0}))]^{2}}.
\end{eqnarray}
The Hubble expansion rate is described by the expression
\begin{equation}\label{206}
   H(t) = \frac{2}{t}\,\frac{1}{3 - \tanh (\lambda (t - t_{0}))}.
\end{equation}
The equations (\ref{205}) - (\ref{206}) reproduce the known expressions for the corresponding quantities
in limiting cases. For $t \rightarrow 0$ and $\lambda t_{0} > 2$, we have 
\begin{equation}\label{207}
   \rho = \frac{3}{32 \pi G t^{2}}, \ w = \frac{1}{3}, \ p = \frac{1}{3} \rho, \ H = \frac{1}{2t}, \ R \sim t^{1/2}.
\end{equation}
For $t \approx t_{0}$, the expressions are the following
\begin{equation}\label{208}
   \rho = \frac{1}{6 \pi G t^{2}}, \ w = 0, \ p = 0, \ H = \frac{2}{3t}, \ R \sim t^{2/3},
\end{equation}
while in the region $t \gg t_{0}$, 
\begin{equation}\label{209}
   \rho = \frac{3}{8 \pi G t^{2}}, \ w = - \frac{1}{3}, \ p = - \frac{1}{3} \rho, \ H = \frac{1}{t}, \ R \sim t
\end{equation}
(cf. Ref. \cite{LL}) and Eq. (\ref{131})). 

Thus, we have a continuous transition from the era when radiation dominates over matter, 
through the era of dust domination, to the epoch when matter in the form of low-velocity cosmic strings dominates.

The transition from the radiation-dominated universe to the universe dominated 
by a perfect gas of low-velocity cosmic strings can be described in terms of simple string-gas model
with the equation of state $p_{s} = w \rho_{s}$, where $w = \frac{2}{3} v_{s}^{2} - \frac{1}{3}$, and
$v_{s}$ is the average velocity of cosmic strings \cite{KT}. At $v_{s} = 1$ the string gas behaves
as relativistic matter, at $v_{s} = \frac{1}{\sqrt{2}}$ it acts as pressure-free matter, and at $v_{s} = 0$
one has a perfect gas of low-velocity cosmic strings.
In such a description, if a gas of low-velocity cosmic strings quickly comes to dominate over relativistic 
and pressure-free matter, it would drastically alter the cosmological evolution of the universe.
In the model of two-component perfect fluid, this problem is removed, since the era with $\rho \sim R^{-2}$ does not start
until the values of the 
radius and mass of the observed part of the universe will become large enough (at least as in the present-day universe).


\begin{thebibliography}{99}
\itemsep -6pt plus 1pt minus 1pt
\bibitem{PDG}K.~Nakamura \textit{et al.} (Particle Data Group), J. Phys. G \textbf{37}, 075021 (2010).

\bibitem{Rio}A.~Riotto, CERN Yellow Report CERN-2010-01, 315 (2010), e-print arXiv:1010.2642 [hep-ph] (2010).

\bibitem{Bia} E.~Bianchi and C.~Rovelli, e-print arXiv:1002.3966 [astro-ph.CO] (2010).

\bibitem{Pe} L.~Perivolaropoulos, e-print arXiv:1104.0539 [astro-ph.CO] (2011); L.~Perivolaropoulos, The Problems of Modern Cosmology, edited by P.M.~Lavrov (Tomsk State Pedagogical University Press, Tomsk, 2009), e-print  arXiv:0811.4684 [astro-ph] (2008); P.~Kroupa \textit{et al.}, Astron. Astrophys. \textbf{523}, A32 (2010), e-print arXiv:1006.1647 [astro-ph.CO] (2010).

\bibitem{Kol} E.W.~Kolb, Astrophys. J. \textbf{344}, 543 (1989).

\bibitem{KT} E.W.~Kolb and M.S.~Turner, \textit{The Early Universe} (Addison-Wesley, Redwood City, 1990).

\bibitem{Whi} G.J.~Whitrow and D.G.~Randall, MNRAS \textbf{111}, 455 (1951).

\bibitem{Mes} D.~Meschini, Found. Sci. \textbf{12}, 277 (2007), e-print arXiv:gr-qc/0601097 (2006); 
C.~Kiefer and B.~Sandhoefer, Beyond the Big Bang, edited by R.~Vaas (Springer, Heidelberg, 2008), e-print arXiv:0804.0672 [gr-qc] (2008).

\bibitem{Ku1} V.V.~Kuzmichev, Ukr. J. Phys. \textbf{43}, 896 (1998); V.V.~Kuzmichev, Phys. Atom. Nucl. \textbf{62}, 708 (1999), e-print arXiv:gr-qc/0002029 (2000); V.V.~Kuzmichev, Phys. Atom. Nucl. \textbf{62}, 1524 (1999), e-print arXiv:gr-qc/0002030 (2000); V.E.~Kuzmichev and V.V.~Kuzmichev, Eur. Phys. J. C \textbf{23}, 337 (2002), e-print arXiv:astro-ph/0111438 (2001).

\bibitem{Ku2} V.E.~Kuzmichev and V.V.~Kuzmichev, Acta Phys. Pol. B \textbf{39}, 979 (2008), e-print arXiv:0712.0464 [gr-qc] (2007); V.E.~Kuzmichev and V.V.~Kuzmichev, Acta Phys. Pol. B \textbf{39}, 2003 (2008), e-print arXiv:0712.0465 [gr-qc] (2007); V.E.~Kuzmichev and V.V.~Kuzmichev, Acta Phys. Pol. B \textbf{40},  2877 (2009), e-print arXiv:0905.4142 [gr-qc] (2009); V.E.~Kuzmichev and V.V.~Kuzmichev, Ukr. J. Phys. \textbf{55},  626
(2010).

\bibitem{Lun} F.~Lund, Phys. Rev. D \textbf{8}, 3247 (1973);
V.G.~Lapchinskii and V.A.~Rubakov, Theor. Math. Phys. \textbf{33}, 1076 (1977); 
F.J.~Tipler, Rep. Prog. Phys. \textbf{68}, 897 (2005).

\bibitem{Tol} R.C.~Tolman, \textit{Relativity, Thermodynamics and Cosmology} (Clarendon Press, Oxford, 1969) \S 139.

\bibitem{Rie} A.G.~Riess \textit{et al.}, Astrophys. J. \textbf{730}, 119 (2011), e-print arXiv:1103.2976 [astro-ph.CO] (2011).

\bibitem{Lar} D.~Larson \textit{et al.}, Astrophys. J. Suppl. \textbf{192}, 16 (2011), e-print arXiv:1001.4635 [astro-ph.CO] (2010); N.~Jarosik \textit{et al.}, Astrophys. J. Suppl. \textbf{192}, 14 (2011), e-print arXiv:1001.4744 [astro-ph.CO] (2010).

\bibitem{Ben} A.~Benoit-Levy and~G. Chardin, e-print arXiv:0903.2446 [astro-ph.CO] (2009); e-print arXiv:0811.2149 [astro-ph] (2008).

\bibitem{Got} J.R.~Gott and M.J.~Rees, MNRAS \textbf{227}, 453 (1987).

\bibitem{Gr} {\O}.~Gr{\o}n, Eur. J. Phys. \textbf{23}, 135 (2002), e-print arXiv:0801.0552 [astro-ph] (2008).

\bibitem{Vil} A.~Vilenkin, Phys. Rep. \textbf{121}, 263 (1985).

\bibitem{Row}M.~Rowan-Robinson, \textit{Cosmology} (Clarendon Press, Oxford, 2004).

\bibitem{Sc1} D.W.~Sciama, MNRAS \textbf{113}, 34 (1953).

\bibitem{Sc2} D.W.~Sciama, \textit{Modern Cosmology} (Cambridge University Press, Cambridge, 1971).

\bibitem{Mach} E.~Mach, \textit{Die Mechanik in ihrer Entwickelung: historisch-kritisch dargestellt} (F.A. Brockhaus, Leipzig, 1897).

\bibitem{Bon} H.~Bondi and J.~Samuel, Phys. Lett. A \textbf{228}, 121 (1997), e-print arXiv:gr-qc/9607009 (1996).

\bibitem{Bar} J.~Barbour, Found. Phys. \textbf{40}, 1263 (2010), e-print arXiv:1007.3368 [gr-qc] (2010).

\bibitem{Whe} J.A.~Wheeler, Gravitation and Relativity, edited by Hong-Yee Chiu and W.F.~Hoffmann (Benjamin, New York, 1964).

\bibitem{Bra} C.H.~Brans and R.H.~Dicke, Phys. Rev. \textbf{124}, 925 (1961).

\bibitem{Hel} M.~Heller, Acta Phys. Pol. B \textbf{1}, 123 (1970).

\bibitem{Di1} P.A.M.~Dirac, Nature \textbf{139}, 323 (1937).

\bibitem{Di2} P.A.M.~Dirac, Pro. Roy. Soc. London A \textbf{333}, 403 (1973).


\bibitem{Wei} S.~Weinberg, \textit{Gravitation and Cosmology} (Wiley, New York, 1972).

\bibitem{Pet} J.P.~Petit, Mod. Phys. Lett. A \textbf{3}, 1527 (1988); J.W.~Moffat, Int. J. Mod. Phys. D \textbf{2}, 351 (1993), e-print arXiv:gr-qc/9211020 (1992); A.~Albrecht and J.~Magueijo, Phys. Rev. D \textbf{59}, 043516 (1999), e-print arXiv:astro-ph/9811018 (1998).

\bibitem{Chi}T.~Chiba, Prog. Theor. Phys. \textbf{126}, 993 (2011), e-print arXiv:1111.0092 [gr-qc] (2011).

\bibitem{Lim} I.~Prigogine, J.~Geheniau, E.~Gunzig and P.~Nardone, Gen. Relativ.
Gravit. \textbf{21}, 767 (1989); J.A.S.~Lima, M.O.~Calvao, and I.~Waga, Cosmology, thermodynamics and matter creation in frontier physics, essays in honor of Jaime Tiomno, edited by S.~MacDowel, H.M.~Nussenzweig, and R.A.~Salmeron (World Scientific, Singapore, 1990) e-print arXiv:0708.3397 [astro-ph] (2007); A.~de~Roany and J.A.~de~Freitas Pacheco, Gen. Relativ. Gravit. \textbf{43}, 61 (2011).

\bibitem{LL}L.D.~Landau, E.M.~Lifshitz \textit{The Classical Theory of Fields. Vol. 2} (Butterworth-Heinemann, Amsterdam, 1975).

\end{thebibliography}
\end{document}